# Neutron lifetime beam anomaly and possible explanation.

A. P. Serebrov,*
*Petersburg Nuclear Physics Institute, NRC "Kurchatov Institute",
188300 Gatchina, Leningrad District, Russia*

## Abstract

The review of experimental measurements of neutron lifetime is presented. Latest measurements with gravitational trap (PNPI NRC KI) and magnetic trap (LANL, USA) confirmed the result obtained by PNPI group in 2005. The results of measurements performed using UCN storing method are in good agreement; however, there is a significant discrepancy at 3.6σ (1% of decay probability) level with beam method experiment. The latest most accurate measurements of the neutron decay asymmetry and neutron lifetime measurements by storage method are in agreement within the Standard Model. This article discusses the possible causes of discrepancy in the measurements of the neutron lifetime. The most probable cause, apparently, is the loss of protons in beam method experiment during storage in a magnetic trap due to charge exchange collisions of protons with the residual gas. The proton becomes neutral and leaves the trap, which leads to a decrease in the number of registered protons, i.e. to a decrease in the probability of neutron decay or to an increase in the measured neutron lifetime.

A. P. Serebrov,* E-mail: serebrov_ap@pnpi.nrcki.ru

## 1. Introduction.
### The history of neutron lifetime measurements.

Neutron lifetime is one of the most important fundamental constants for the weak interaction theory and cosmology. Neutron has the longest lifetime among the unstable elementary particles; its lifetime is ~880 seconds. It is the great length of lifetime, i.e. very small decay probability, to be the reason that parameter is very hard to measure. For example, in a cold neutron beam at 1 m distance only one of a million neutrons passing through experimental setup occurs to decay. However, there is an alternative way to measure neutron lifetime using ultracold neutrons (UCN). These neutrons have very low kinetic energy, they reflect off the walls of material traps and magnetic traps with magnetic field gradients at the wall. The idea of the experiment is to store neutrons into the trap and observe their decay. The loss probability in the trap can be decreased to 1-2% of the neutron decay probability, applying the cryogenic material traps [1, 2] and even lower losses are achievable with magnetic traps [3, 4, 5, 6]. That means neutrons can be stored in traps and neutron lifetime can be measured almost directly, introducing small corrections for UCN losses in the trap.

The history of neutron lifetime measurements covers the significant period of time starting from the first experiments in 70s at neutron beams [7, 8]. Since then the accuracy of measurements has increased over an order of magnitude, and significant progress was achieved using UCN. One must, however, recall the pioneering work by A Snell (USA, 1950), J Robson (Canada, 1950), and P E Spivak (USSR, 1955) performed with neutron beams.

However, the progress in UCN method was not as certain as it might seem. The first experiments with UCN storing lacked accuracy due to small UCN density into a trap [9]. Accuracy of the experiments increased after the UCN sources with high intensity were created in Gatchina [10] and Grenoble [11]. The significant success was achieved by using fluorine containing oil (fomblin), where the Hydrogen atoms replaced by fluorine [12, 13]. However, the probability of UCN losses in those experiments was ~30% [12] and ~13% [13] of neutron decay probability. The experimental problem was the extrapolation of UCN trap storage time to neutron lifetime, performing measurements with various collision frequencies using various trap geometry. Extrapolation range was about ~200 $s$ [12] and ~100 $s$ [13], hence achieving 1 $s$ accuracy for extrapolation was an extremely difficult task.

Besides, the effect of low energy upscattering was discovered and it leads to systematic effect in neutron lifetime measurements [14, 15, 16, 17]. UCN measurements of neutron lifetime were significantly improved by applying an open-topped cryogenic trap where neutrons are trapped by Earth gravity [18]. Using low temperatures the effects of inelastic scattering and low energy upscattering were suppressed and loss probability at walls became about 1÷2% of neutron decay probability. Here the extrapolation range becomes only 5 ÷ 10 s. That way the accuracy of 1 $s$ for neutron lifetime is achievable.

Within the experiment carried out in 2004 in ILL by the collaboration of PNPI and JINR [1] was obtained the neutron lifetime $878.5 \pm 0.7 \pm 0.3\ s$, here the first error is statistical and the second is systematical. The result of experiment carried out in Gatchina [18] with similar trap was in good agreement within accuracy and the difference was less than $2\sigma$. Neutron lifetime value in PDG 2006 was $885.7 \pm 0.8\ s$. The discrepancy between the result of the new experiment carried out in 2005 [1] and PDG value was $6.5\sigma$ and caused a wide discussion with significant mistrust in that discrepancy. However, in two years in the first experiments with magnetic trap with permanent magnets [3] the result was confirmed and the measured lifetime was $878.2 \pm 1.9\ s$. This result was presented at the VII International Conference, Ultracold and Cold Neutrons. Physics and Sources [19] in 2009, and later it was published [5].

In 2010 in the experiment MAMBO II [20] was obtained the result $880.7 \pm 1.8\ s$. In 2012 the results of the experiments with room temperature fomblin [12, 13] were corrected to be $882.5 \pm 2.1\ s$ [21] and $881.6 \pm 2.1\ s$ [22]. Finally in 2015 the new experiment was carried out by scientific group led by V. I. Morozov and the result was $880.2 \pm 1.2\ s$ [23]. Back in 2010 our scientific group (PNPI, Gatchina) have developed a project of the experiment with big gravitational trap to check the result of our experiment carried out in 2005 [24]. This experiment with big gravitational trap was completed by PNPI NRC KI-ILL-RAL collaboration in 2017 [2, 25] and it obtained the result $881.5 \pm 0.7 \pm 0.6\ s$, so within $2\sigma$ both results turned out to be consistent. In the same year 2017 the result of Los Alamos national laboratory with the magnetic trap for UCN was published. [6]: $877.7 \pm 0.7 \pm 0.3\ s$.

Summarizing all those results we can conclude that the result obtained in 2005 is confirmed by the experiments with UCN storing. The historical diagram of measurements starting 1990 is shown in fig. 1.

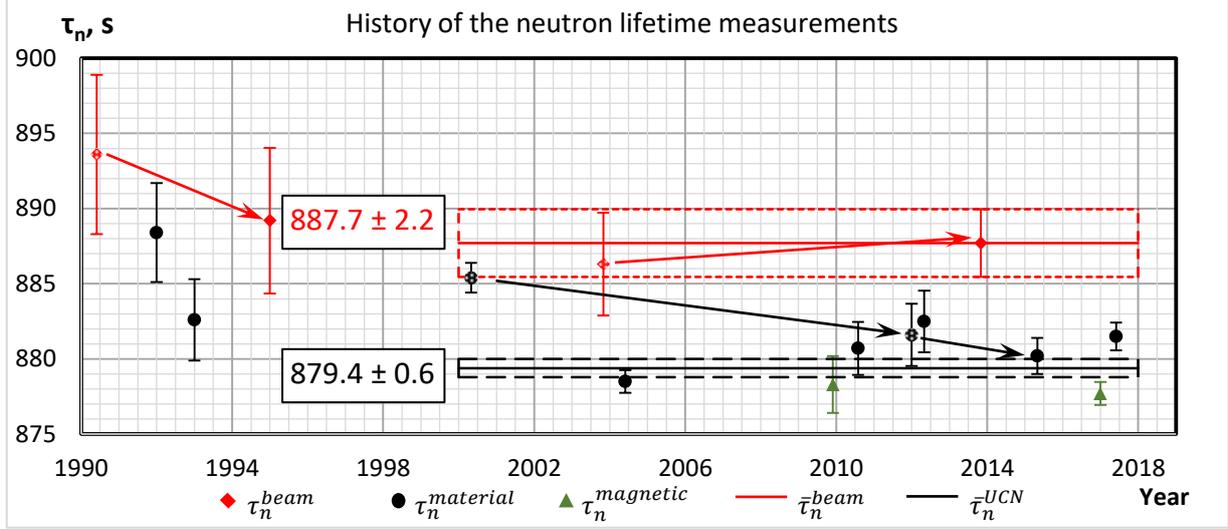

Fig. 1. Experimental results of neutron lifetime measurements from 1990, the discrepancy of the results obtained in 2005 [1] and 2000 [13], the correction of liquid fomblin experiment [21, 26] and the new experiment [23], finally, the new results of 2017 [6] and [2]. $\tau_n^{beam}$ — beam experiments, $\tau_n^{material}$ — material traps, $\tau_n^{magnetic}$ — magnetic traps, $\bar{\tau}_n^{beam}$ — beam average, $\bar{\tau}_n^{UCN}$ — UCN storage average. Dashed lines represent the corridor of errors of the mean value. The shaded points indicate the results that later were corrected or revised by the authors.

In fig. 2 the diagram of neutron lifetime measurements is presented starting from 2005. On the left one can see the results of the experiments with UCN storing in material and magnetic traps. From the data one can conclude that the results of storing experiments are consistent within two standard deviations. On the right are the results from neutron beams with proton trap, which significantly differs [27, 28]. In the table are listed the results of the experiments with statistical and systematic errors and also the total error calculated as a squared sum of errors.

The discrepancy between beam [27, 28] and UCN storing experiments is 3.6σ if we use quadratic addition and 2.7σ if we use linear addition. In any case it is a noticeable discrepancy [29], and it is sometimes called "neutron anomaly" [30, 31].

If the result obtained with USNs alone are averaged then one obtains $\tau_n = 879.4 \pm 0.6\ s$, while the value of $\chi^2$ is reduced from c 1.9 to 1.5.

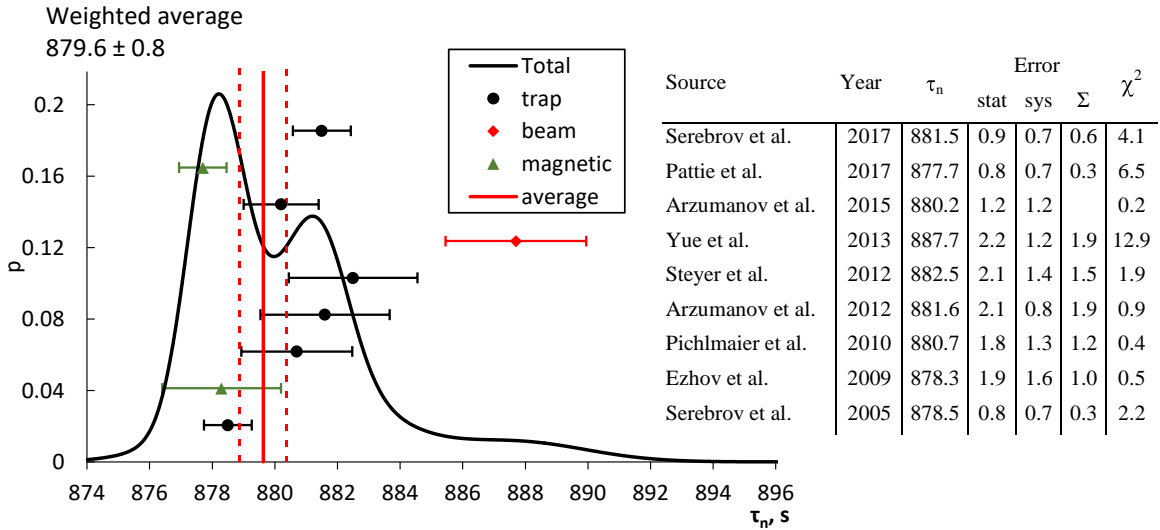

| Source | Year | $\tau_n$ | stat | sys | Σ | $\chi^2$ |
|---|---|---|---|---|---|---|
| Serebrov et al. | 2017 | 881.5 | 0.9 | 0.7 | 0.6 | 4.1 |
| Pattie et al. | 2017 | 877.7 | 0.8 | 0.7 | 0.3 | 6.5 |
| Arzumanov et al. | 2015 | 880.2 | 1.2 | 1.2 | | 0.2 |
| Yue et al. | 2013 | 887.7 | 2.2 | 1.2 | 1.9 | 12.9 |
| Steyer et al. | 2012 | 882.5 | 2.1 | 1.4 | 1.5 | 1.9 |
| Arzumanov et al. | 2012 | 881.6 | 2.1 | 0.8 | 1.9 | 0.9 |
| Pichlmaier et al. | 2010 | 880.7 | 1.8 | 1.3 | 1.2 | 0.4 |
| Ezhov et al. | 2009 | 878.3 | 1.9 | 1.6 | 1.0 | 0.5 |
| Serebrov et al. | 2005 | 878.5 | 0.8 | 0.7 | 0.3 | 2.2 |

Fig. 2. Neutron lifetime measurements diagram from 2005 in experiments with storing UCN in magnetic and material traps, and also in the neutron beam experiment with proton trap to detect neutron decay protons. p — probability density. The uncertainty of the mean value represented by red dashed lines are scaled by $\sqrt{\chi^2} = 1.9$.

## 2. Analysis of the discrepancy between beam and UCN storing measurement methods

First of all, we should analyze the essential difference in measurement procedures with UCN and beam methods.

The beam experiment is based on the expression:
$$\Delta N_p = \lambda N_n \Delta t \quad (1)$$

Where $\Delta N_p$ is a number of registered products of neutron decay (protons or electrons) obtained while beam pass the setup, $N_n$ is a number of neutrons passed the setup, $\Delta t$ neutron time of flight and $\lambda = 1/\tau_n$ is neutron decay probability. Here is assumed that there is only one decay channel in p, e, $\tilde{\nu}$. The probability of decay into hydrogen atom is negligible and estimated to be $3.9 \cdot 10^{-4}\%$.

The main difficulty in beam experiment is absolute measurements of values in equation (1), and also an efficiency of proton and electron registration.

The experiment with UCN storing method is based on measurements of time dependence:
$$N_n(t) = N_n(0) e^{-t/\tau_{storage}} \quad (2)$$

Where $N_{n,p,e}(t)$ — number of neutrons in a trap in moment t, which can be measured using neutron detector for several time intervals, or it is the rate of neutron decays in the trap, which can be measured using electron or proton detector, $\tau_{storage}^{-1}$ — storage probability in the trap:
$$\tau_{storage}^{-1} = \tau_n^{-1} + \tau_{loss}^{-1} \quad (3)$$

Main difficulty in UCN method is a precise measurement of UCN loss probability in the trap — $\tau_{loss}^{-1}$. Losses in the trap are determined by frequency of collisions with walls of the trap and UCN interaction with residual gas:
$$\tau_{loss}^{-1} = \eta \gamma(E) + \tau_{vac}^{-1} \quad (4)$$

Where $\eta$ — loss factor, which does not depend on UCN energy, $\gamma(E)$ — effective collision frequency, which depends on UCN energy and trap size, $\tau_{vac}^{-1}$ is probability of UCN loss in interaction with residual gas molecule.

In experiments [1, 2, 9, 12, 18, 20, 21] is measured the dependence of $\tau_{loss}^{-1}$ on collision frequency and performed an extrapolation of $\tau_{storage}^{-1}$ to $\tau_n^{-1}$.

In experiments [13, 22, 23] collision frequency is measured by registration of neutrons after inelastic interaction with trap walls using $^3He$ thermal neutrons detectors installed outside the trap.

In experiments [1, 2, 18] with gravitational UCN trap at low temperature (~100K) loss factor is rather small and the extrapolation $\tau_{storage}^{-1}$ to $\tau_n^{-1}$ is about $10 - 15\ s$, hence accuracy of about $\pm 1\ s$ is achieveable.

Finally, in experiments with UCN storing in magnetic traps [19, 6] UCN losses during storing should be zero in assumpton of zero depolarisation in strong magnetic fields. The results of two independent measurements [19] and [6] are in good agreement.

In summary, there is a good agreement in results of seven experiments with UCN storing, including consistency of measurements with storing UCN in material and magnetic traps. One should accept, that the result of neutron lifetime $(879.4 \pm 0.6\ s)$, obtained in series of seven experiments with various methods is reliable.

Beam experiment [28] is the one most accurate of beam experiments, its accuracy override previous beam experiments. The descrepancy between one beam experiments and the series of UCN storing experiments should not be called "neutron anomaly" yet, at least, one have to repeat the experiment [28] and carry out independent beam experiments.

Naturally, in current situation of searching for "new physics" the interest to that problem is totally understandable. Any discrepancy at $3\sigma$ level becomes a matter of discusson. So we would like to look through and list here the ideas discussed before and under discussion now, which aims to explain the measurement discrepancy. Most of assumptions were about presence of unaccounted losses in experiments with UCN storing.

**1.** One of the most popular hypotheses is the so-called "small heating" during storage of UCNs in traps. Recently, a paper [32] was published in which even the effect of the Earth's rotation on the storage of UCNs in traps is considered. Indeed, due to the rotation of the trap and due to the interaction of UCNs with the walls of the trap, a slow broadening of the spectrum of stored neutrons will occur (heating and cooling). Due to the increase in energy, a neutron can leave the trap. In paper [32] it is proposed to take this effect into account in experiments on the storage of UCNs when it comes to accuracy better than 1%. In this regard, it should be noted that in an experiment with a large gravitational trap, the effect of "heating" UCNs during storage in a trap is under control. "Heated" neutrons would leap out of the trap and would be revealed by the detector during a long storage interval of 1600 s. An experimental estimate of the upper limit of such an effect is less than one second. In addition, this effect is compensated by extrapolation to the zero collision frequency, i.e. when extrapolating to neutron lifetime.

**2.** When in 2005 was obtained result $878.5 \pm 0.7 \pm 0.3\ s$ [1] with $6.5\sigma$ deviation from PDG value one of the assumptions was oscillation $n \rightarrow n'$ (neutron – mirror neutron [33]). The idea of this proposal must be clarified.

Our world is left for weak interaction and the problem of global symmetry restoration is under consideration for long period [34]. Considering global symmetry restoration, one can assume that dark matter world is right corresponding to space inversion. In the simplest model of "mirror Standard Model" mirror neutron $(n')$ is a dark matter particle having same mass as neutron but with opposite value of magnetic momentum and very small interaction with standard matter constant, but same gravitational interaction. In that model transitions $n \rightarrow n'$ are possible in absence of magnetic fields (both usual magnetic fields and mirror magnetic fields of dark matter). After the transition mirror neutron leaves the trap, because it has almost zero interaction with matter. That means neutron β-decay probability in storage experiments would be overestimated. An assumption of possibility of $n \rightarrow n'$ oscillation was considered in work [33] in 2006. Experimental searches for $n \rightarrow n'$ oscillation were performed in works [35, 36, 37, 38]. The most accurate limit on $n \rightarrow n'$ oscillation was obtained in work [36]. There it was shown that oscillation period is higher than $414\ s$ (90% C.L.) or the oscillation probability is less than $2.4 \cdot 10^{-3} \, s^{-1}$ in absence of magnetic field. In 2009 the result of the $n \to n'$ oscillation period upper limit was increased to $448 \, s$ (90% C.L.) [37]. Thus, the $n \to n'$ oscillations were not observed. Those researches are undoubtedly very interesting. However, it should be noticed that neutron oscillation cannot explain the discrepancy between beam and UCN experiments.

The reason is that even if $n \to n'$ oscillation exists it is significantly suppressed by Earth magnetic field. Magnetic field affects both magnitude and frequency of the oscillation, and even Earth magnetic field is enough to make oscillations very short, much shorter than neutron path between collisions. Therefore, the effect of UCN leaving the trap through mirror matter component is proportional to collision frequency and it is excluded in extrapolation to zero collision frequency. Hence, the idea of $n \to n'$ oscillations cannot explain discrepancy of two measurement methods (beam and UCN).

**3.** In standard scheme of neutron decay, three modes are considered, though it mostly determines by the decay where proton occurs in final state and only about 1% has both $\gamma$-quantum and proton.

$n \to p + e^- + \bar{\nu}_e$           100%
$n \to p + e^- + \bar{\nu}_e + \gamma$     $(9.2 \pm 0.7) \cdot 10^{-3}$      [39]
$n \to H + \bar{\nu}_e$               $3.9 \cdot 10^{-6}$              [40]

The $\gamma$ quantum appears as a result of bremsstrahlung process of decay electron and its energy depends on electron energy by $E_\beta^{-1}$. Relative probability of this process is about 1%, but it is automatically taken into account in experiment [28], because it has the proton in the final state.

The process, which is more suitable to concern neutron anomaly, is the neutron decay into hydrogen atom, which cannot be held in electro-magnetic trap in experiment [28] but it has very small relative probability of about $3.9 \cdot 10^{-4}$% [40]. Yet 1% relative probability is required to explain neutron anomaly. However, it should be noted that it would be interesting to calculate the correction for the probability of the formation of a hydrogen atom when the neutron decays in a sufficiently strong magnetic field of $4.6 \, T$. Naturally, such a strong magnetic field cannot affect the total probability of neutron decay, but the possibility of the influence of the magnetic field on the formation the hydrogen atom in the final state should be evaluated. This estimation was performed by E.G. Druckaryov (PNPI NRC KI), and revealed that the magnetic field changes the probability of the formation of a hydrogen atom negligibly.

**4.** Recently an interesting explanation of the neutron decay anomaly was published in work [31]. It is based on introducing additional decay channel into dark matter in final state. Assuming those particles are stable in final state then they can be the dark matter particles with mass close to neutron mass Regarding the ideas discussed above this transition into dark matter is very similar to transition into mirror neutron - dark matter particle with mass close to neutron mass. It should be noticed that in the dark matter model the interaction of dark matter with baryons is assumed. In this scenario a monoenergetic photon in energy range $0.782 - 1.664 \, MeV$ is yielded in neutron lifetime experiment with 1% branching [31]. That is very important and reveals that experimental test is possible. That experimental test [41] was performed almost right after the publication [31]. At $4\sigma$ confidence level monochromatic γ-quanta were not observed.

**5.** As a development of the mirror dark matter idea in paper [42] was considered a scheme with the mass of a mirror neutron being less than the mass of a standard neutron. The article is devoted to an attempt to connect the "neutron anomaly" with the so-called "reactor antineutrino anomaly", which means the deficit of the measured antineutrino flux from the reactor with respect to the calculated flux. The problem is actively discussed at the neutrino conferences, and the experiments dedicated to the search of sterile neutrino i.e. transition to the dark matter in the neutrino sector are being performed. In paper [42] two so-called anomaly was discussed: the neutrino and reactor anomalies. Both antineutrino and neutron anomalies are at confidence level of $\sim 3\sigma$ («antineutrino deficit» is $6.6 \pm 2.4$%, and «neutron anomaly» is $1.0 \pm 0.3$%). The peculiarity of the proposal of this article is that both anomalies can be explained by one phenomenon of oscillations in the baryon sector between the neutron and the dark matter neutron $n \to n'$ with mass $m_{n'}$ slightly less than $m_n$ of ordinary neutron. The mass difference $m_n - m_{n'}$) can be compensated by the binding energy in the nucleus, and the transitions $n \to n'$ will be amplified. The calculations of the proposed model require one free parameter: the mass difference $m_n - m_{n'}$. If probability $n \to n'$ 'of oscillations for a free neutron is normalized to the "neutron anomaly (1%)", than upon achieving in calculations an explanation of the "neutrino anomaly" (6.6%) one can determine the mass difference $m_n - m_{n'}$ and thus the mass of the dark matter neutron. Preliminary estimates imply that suitable mass difference is $m_n - m_{n'} \approx 3 \, MeV$. However, the analysis of the data of cumulative isotope yields of fission products was carried out, which does not confirm the possibility of an additional decay channel with the release of dark matter neutrons with a mass difference $m_n - m_{n'} \approx 3 \, MeV$. The result of the analysis is the conclusion that, for mirror neutrons, the region of mass difference $m_n - m_{n'} \geq 3 \, MeV$ is closed.

**6.** In the recent paper [43] the scheme of mirror dark matter is considered with $m_n - m_{n'} \approx 10^{-7} \, eV$. The assumption is made that when neutron passes through the magnetic field of a solenoid in an experiment [27, 28], then the mass difference $m_n - m_{n'}$, is compensated due to the binding energy in the magnetic field due to the magnetic moment of the neutron. The $n \to n'$ 'transitions are amplified, and the fraction of standard decays with the appearance of a proton decreases by 1%. Such an assumption can be investigated experimentally [28], by varying the magnetic field, and also in a new beam experiment [44] with a magnetic field 5 times smaller, which is currently under preparation

**7.** In attempts to find the difference in the design of the beam experiment and experiments with the storage of UCNs, it can be noted that in the beam experiment, neutron decay is observed for cold neutrons, and not for UCNs. It is not possible to point to any physical reason following from that. However, an actual difference exists in the time of decay observation. In a cold neutron beam, the decay process is observed in the interval $10^{-3} \, s$ after the last neutron interaction (collision with the neutron guide wall). In the experiment with UCNs, the average flight time between collisions is $0.3 \, s$, and the measure-ment interval of the decay exponent is generally ~$10^3$ s. Comparing these intervals, one can raise the question of how rigorously the exponential law is satisfied. Deviation from the exponential law can occur when there are levels in the initial state or several decay modes [45]. Finally, one can recall the Zeno quantum paradox [46] according to which all unstable states freeze at $t = 0$ as well as the Zeno quantum effect, according to which the decay probability can change if you measure it often enough. The Zeno paradox can be associated with measurements on the beam, because $t = 10^{-3} s$, and the neutron lifetime is six orders of magnitude larger. With measurements with UCN one can associated the quantum Zeno effect, since frequent collisions of a neutron with the walls of the trap correspond to acts of measuring the neutron stability at each moment in time. There are $10^4$ of such measurements. . Whether the Zeno quantum paradox and the Zeno quantum effect can be related to the problem under consideration remains to be discovered. According to estimates given in Refs [45, 47], the time scale at which the decay law of unstable particles can differ from exponential is far beyond the characteristic times that one has to deal with in experiments on measuring the neutron lifetime.

In summary, the conclusion can be made that there still exists no clear physical idea, which would be able to explain the observed contradiction. It is possible that the most probable answer is the systematic error presenting in the beam experiment.

## 3. Measurements of neutron decay asymmetry and Standard Model test

Consider in details the researches of neutron decay including measurements of asymmetry of β-decay and tests of SM. It is well-known, the matrix element $V_{ud}$ that Cabibbo-Kobayashi-Maskawa (CKM) matrix:

$$\begin{pmatrix} d' \\ s' \\ b' \end{pmatrix} = \begin{pmatrix} V_{ud} & V_{us} & V_{ub} \\ V_{cd} & V_{cs} & V_{cb} \\ V_{td} & V_{ts} & V_{tb} \end{pmatrix} \begin{pmatrix} d \\ s \\ b \end{pmatrix} \quad (5)$$

can be determined from β-decay by measuring neutron lifetime and decay asymmetry (fig. 3) and the result can be compared with other methods of $V_{ud}$ calculations. The formula relating neutron half-life period ($\tau_{1/2}$) with the $V_{ud}$ element is:

$$f\tau_{1/2}(1 + \delta_R')(1 + \Delta_R) = \frac{K}{|V_{ud}|^2 G_F^2 (1 + 3\lambda^2)} \quad (6)$$

where $f = 1.6886$ — is a phase space factor; $\delta_R' = 1.466 \times 10^{-2}$ is a model-independent radiative correction calculated with precision of $9 \cdot 10^{-5}$, $\Delta_R = 2.40 \times 10^{-2}$ — model-dependent internal radiative correction, calculated with precision of $8 \cdot 10^{-4}$, $G_F$ — is the Fermi weak interaction coupling constant determined from μ-decay; $K = \hbar(2\pi^3 \ln 2)(\hbar c)^6/(m_e c^2)^5$, $\lambda = G_A/G_V$ — is the ratio of the axial-vector and vector weak interaction coupling constants, determined experimentally from measurements of angular correlation coefficients in the neutron β-decay. In the experiment of measuring the beta-decay asymmetry following quantity is determined:

$$A_0 = -2 \frac{\lambda(\lambda + 1)}{(1 + 3\lambda^2)} \quad (7)$$

Considering that $G_V = V_{ud} G_F$ in fig. 3 from equation (6) we obtain an ellipse, and from equation (7), the curve, that crosses it, and the intersection point permits us to determine $V_{ud}$ element:

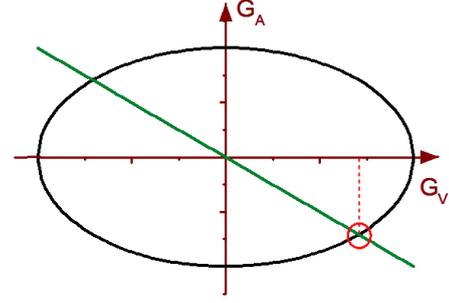

Fig. 3. Determination of the $V_{ud}$ matrix element from the neutron decay data.

Formula (6) for the $V_{ud}$ element can be rewritten as follows [48]:

$$|V_{ud}|^2 = \frac{4908.7 \pm 1.9 \, s}{\tau_n(1 + 3\lambda^2)} \quad (8)$$

In fig. 4 are presented the results of a test of data on the neutron β-decay in order to determine $V_{ud}$, making use of the ration of the axial an vector weak interaction coupling constants ($G_A/G_V = \lambda$) based on the most precise measurements of the electron decay asymmetry [49].

The intersection of data for $\tau_n$ and $\lambda = G_A/G_V$ yields a value of $V_{ud}$ from the neutron decay that can be compared with the value $V_{ud}$ from super-allowed $0^+ \to 0^+$ nuclear transitions and with the value of $V_{ud}$ from the unitarity of CKM matrix ($V_{ud}^2 + V_{us}^2 + V_{ub}^2 = 1$).

As one can see in fig. 4 the test of the Standard Model is passed successfully only if neutron lifetime data used come from UCN storage experiments.

Therefore for the tasks involving elementary particle physics, astrophysics, cosmology and neutrino physics it's preferable to use value of $879.4 \pm 0.6 \, s$ coming from UCN experiments while beam experiments should be amended.

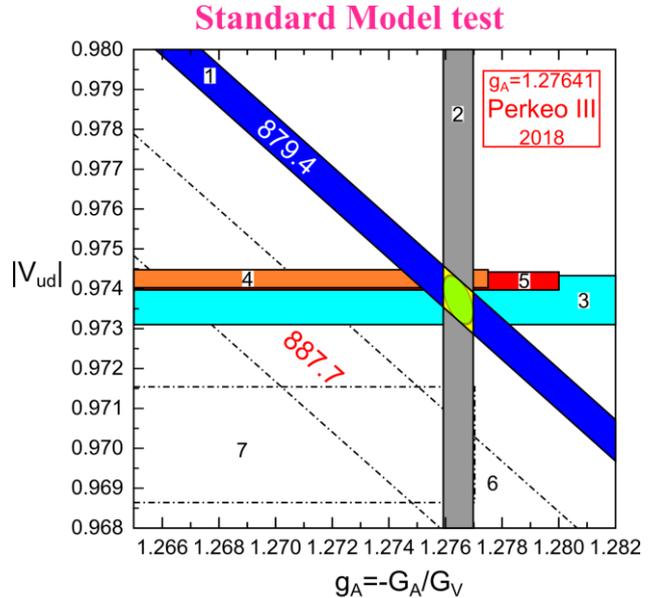

Fig. 4. Dependence of the CKM matrix element $V_{ud}$ on the values of the neutron lifetime and the axial coupling constants $g_A$. (1) neutron lifetime storage experiments measured with UCN ($879.4 \pm 0.6 \, s$); (2) neutron β-decay asymmetry, PERKEO III; (3) neutron β-decay, storage experiments + PERKEO III; (4) unitarity; (5) $0^+ \to 0^+$ nuclear transitions, (6) neutron lifetime Yue 2013; (7) neutron β-decay, Yue 2013 + PERKEO II.

In addition we note that nowadays the accuracy of UCN experiments in measuring neutron lifetime has reached the level of $7 \cdot 10^{-4}$. For the electron asymmetry in the PERKEO II experiment precision of $4.2 \cdot 10^{-3}$ has been reached [49]. The accuracy of results obtained in PERKEO III experiment and presented to the Workshop on Particle Physics at Neutron sources (PPNS-2018) was 2.5 times better and was in agreement with the previous result. Thus, measurements of decay asymmetry exhibit quite decisive agreement. In general, one can conclude that the accuracy of experiments is already approaching the theoretical accuracy related to the calculation of radiative corrections.

4. **Storage of the protons in a magnetic trap and the phenomenon of proton charge exchange in the interaction with the residual gas**

The overestimated result for the lifetime in a beam experiment can be explained by the decrease of the proton storage time due to interaction with the residual gas. In paper [50] it was pointed out that the pressure of the residual gas was $10^{-9}$ mbar, but it was measured near the ion pump at room temperature. It was also noted that the pressure in the trap should be much less, because it is at the temperature of liquid helium. However, the effect of the interaction of protons with the residual gas is determined by the density of the particles of the residual gas, and not by pressure. Based on the equality of particle flows from the warm part of the trap to the cold part and vice versa, we can write the condition for the equality of flows in the equilibrium state $n_1 v_1 = n_2 v_2$, then the particle density in the cold part of the trap is $n_2 = n_1 \sqrt{T_1/T_2}$, where $n_2$ — is the particle density in the cold part and $n_1$ — in the warm part of the trap. $T_1, T_2, v_1, v_2$ — are the corresponding temperatures and mean velocities of molecules. At $T_1 = 300\ K$ and $T_2 = 4\ K$ the particle density in cold part exceeds the particle density in warm part by almost an order of magnitude.

The typical proton charge exchange cross section for $CO, CO_2, O_2, H, H_2O, CH_3, CH_4$ is $2 \div 3 \cdot 10^{-15}\ cm^2$, and the number of particles in the cold part of the trap is $\sim 2 \cdot 10^8\ cm^{-3}$. If the pressure in the warm part of the trap is $10^{-9}\ mbar$ then:

$$n\sigma = 2.5 \cdot 10^{-15} \cdot 2 \cdot 10^8 = 5 \cdot 10^{-7} cm^{-1} \quad (9)$$

For the mean proton energy $500\ eV$ their velocity is:

$$v = 1.38 \cdot 10^{-6} \sqrt{E[eV]} = 3.2 \cdot 10^7 cm/s \quad (10)$$

With an average proton storage time in the trap of $5\ ms$ its traveled distance is $3.1 \cdot 10^7 \cdot 5 \cdot 10^{-3} = 1.5 \cdot 10^5\ cm$. Then $n\sigma l = 5 \cdot 10^{-7} \cdot 1.5 \cdot 10^5 = 7.5 \cdot 10^{-2}$, i.e. loss factor is $\sim 7\%$. As a result of the interaction, the proton becomes neutral and is not registered. The proton charge exchange on the hydrogen atom is not critical, because as a result, another charged low-energy proton appears anyway. It can also be detected after acceleration by an electric field of $30\ kV$. The most dangerous effect is charge exchange at $H_2O, O_2, CH_3, CH_4$, because although a charged molecule is accelerated by the same electric field of 30 kV, it is not detected due to losses in the dead layer of a silicon detector. Thus, the systematic error in measuring the neutron lifetime may reach 7%. Apparently, this is an upper estimate of the possible systematic error of the beam experiment [28]. When taking into account the effect of increasing the density of particles in a cold trap $\left(\sqrt{T_1/T_2} = 8\right)$ entirely.

If the residual gas pressure in the beam experiment was smaller than $10^{-9}\ mbar$, for example by an order of magnitude, than the effect caused by that reduces to 1% which is still in the anomaly range. Note that the effect of the residual gas was already considered in [51], but the effect of cold vacuum was not taken into account.

5. **Conclusion**

The analysis of possible physical effects in the experiment on measuring the neutron lifetime in the beam was carried out, but no explanation was found. Therefore, we had to turn to the analysis of possible systematic errors of the experiment. The most probable cause, apparently, is the loss of protons in the beam experiment during storage in a magnetic trap. Due to proton losses in a magnetic trap, an error with an overestimation of the neutron lifetime might occur.

It is important to note that to suppress the discussed effect it is advisable to use a warm trap, i.e. solenoid with a cavity at room temperature.